\documentclass[final,10pt,times, twocolumn]{elsarticle}

\usepackage{amssymb}
\usepackage{amsmath}

\usepackage{color}

\newcommand{\upd}{\mathrm{d}}
\newcommand{\changes}[1]{{#1}}

\journal{~}

\begin{document}

\begin{frontmatter}

\title{Nonlinear Dielectric Decrement of Electrolyte Solutions: \\
an Effective
Medium Approach}

\author[inst1]{Yasuya Nakayama}
\ead{nakayama@chem-eng.kyushu-u.ac.jp}

\affiliation[inst1]{organization={Department of Chemical Engineering, Kyushu University},
            addressline={Nishi-ku}, 
            city={Fukuoka},
            postcode={819-0395}, 
            country={Japan}}

\begin{abstract}
\paragraph{Hypothesis}
The dielectric constant of an electrolyte solution, \changes{which determines electrostatic interactions between colloids and interfaces,} depends 
nonlinearly on the salinity and also on the type of salt.
The linear decrement at dilute solutions is due to the reduced 
polarizability in the hydration shell around an ion. 
However, the full hydration volume cannot explain the 
experimental solubility, which indicates the hydration volume 
should decrease at high salinity. 
Volume reduction of the hydration shell is supposed to weaken 
dielectric decrement and thus should be relevant to 
the nonlinear decrement.
\paragraph{Simulations}
According to the effective medium theory for the permittivity of 
heterogeneous media, we derive an equation which relates the 
dielectric constant with the dielectric cavities created by the 
hydrated cations and anions, and the effect of partial 
dehydration at high salinity is taken into account.  
\paragraph{Findings}
Analysis of experiments on monovalent electrolytes suggests that 
weakened dielectric decrement at high salinity
originates primarily from the partial dehydration.  
Furthermore, the onset volume fraction of the partial dehydration 
is found to be salt-specific, and is correlated with the 
solvation free energy. 
Our results suggest that while the 
reduced polarizability of the hydration shell determines the 
linear dielectric decrement at low salinity, ion-specific 
tendency of dehydration is responsible for nonlinear dielectric 
decrement at high salinity.
\end{abstract}


\end{frontmatter}

\section{Introduction}
Electrostatic interactions between charged objects such as ions,
colloids, and interfaces immersed in aqueous electrolyte
solutions play an important role in electrochemical processes,
biological systems, and transport
phenomena~\cite{Stone2004ENGINEERING,2006Soft,Bazant2009Towards,Israelachvili2011Intermolecular,saha2022shielding,gao2023effects}.
The phenomenon of dielectric decrement is experimentally well-known and describes the decrease of the dielectric constant of electrolyte
solutions when their salinity increases~\cite{Hasted1948Dielectric,Barthel1995Electrolyte,BenYaakov2011Dielectric}.
The dielectric decrement originates from two effects: a dielectric cavity
created by the ions themselves, and the presence of a hydration
shell around the ions.
For small and
simple ions, such as halides, the ionic cavity hole effect does not have a
substantial contribution to the net dielectric decrement. A more
significant effect is due to the hydration shell
formed by polar water molecules in the immediate proximity to an
ion~\cite{BenYaakov2011Dielectric}. In this layer, the polar water
molecules are largely oriented along the electrostatic field lines created by
the ion, reducing the overall orientational polarizability of the
aqueous solution, and leads to a rather pronounced dielectric decrement.
In relatively dilute solutions (typically for salt concentration, \(n_{b}\), less
than 1.5\,M), the dielectric decrement is linear,
\(\varepsilon(n_{b})=\varepsilon_{w}-\gamma_{b}n_{b}\), where
\(\varepsilon_{w}\approx 80\) is the dielectric constant of pure water and
\(\gamma_{b}\) is the coefficient (in units of M\(^{-1}\)) of the linear term.
At higher \(n_{b}\) values, \(n_{b}\gtrsim 1.5\)\,M, 
however, the dielectric decrement shows a
more complex nonlinear
dependence~\cite{Barthel1995Electrolyte,Glueckauf1964Bulk,Wei1990Dielectric,Wei1992Ion,Levy2012Dielectric,Levy2013Dipolar},
which levels off to a smaller decrement than the linear one.  Not
only the value of the linear coefficient \(\gamma_{b}\), but also the nonlinear
behavior of \(\varepsilon(n_{b})\) is salt-specific, and is related to
the Hofmeister series~\cite{BenYaakov2011Dielectric,Kunz2009Specific}.

Different phenomena in electrolyte solutions 
\changes{both in bulk and in contact with interfaces} stem from the variation of
dielectric constant as \(\varepsilon\) changes the scale of the
electrostatic interactions.
Due to the enhanced electrostatic repulsion caused by the dielectric
decrement, the distribution of counter-ions in the {\itshape electric
double layer}~(EDL) is
broaden~\cite{BenYaakov2011Dielectric,Biesheuvel2005Volume} close to
charged surfaces.  
This makes the electrostatic interaction between charged surfaces more
long-ranged~\cite{BenYaakov2011Dielectric,LopezGarcia2014Influence}.
In this
regard, depletion of polyelectrolytes as driven by dielectric decrement
near oppositely charged surfaces can occur~\cite{Biesheuvel2005Volume}.  For
high-voltage electrodes, the dielectrophoretic saturation of
counter-ions can occur
in the EDL~\cite{Hatlo2012Electric,Nakayama2015Differential},
and presumably is related to layering of
counter-ions in the microscopic scale~\cite{Gillespie2015Review}.
Furthermore, the relation between this dielectrophoretic saturation and
the peaks in the surface differential capacitance as a function of the
surface voltage was addressed by several
authors~\cite{Hatlo2012Electric,Nakayama2015Differential,Bikerman1942XXXIX}.
\changes{Basically, the electrostatic interaction becomes stronger with 
decreasing dielectric constant. Therefore, the enhanced 
electrostatic repulsion between the counterions within EDL 
induces an effect similar to the ion finite size effect~\cite{Nakayama2015Differential}.
Depending on ion type, dominance between ion size and dielectrophoretic repulsion differs. Therefore both effects should be properly included in EDL modelling.
Moreover, since the electrostatic field increases with decreasing dielectric constant, the strength of dielectric decrement also affects the ion-ion correlation.
The nonlinear dielectric decrement at high ion concentrations, which also depends on the ion type, modulates these coupling effects between ion finite size effects, dielectric decrement, and ion-ion correlations.
These effects are most pronounced at high salt concentrations and/or high ionic concentrations near interfaces.
The EDL structure at hard~\cite{gupta2018electrical} and soft interfaces~\cite{lesniewska2023electrostatics} has been discussed using a modified Poisson--Boltzmann model that includes ion 
size, dielectric decrement and ion-ion correlation effects, by assuming linear decrement and concentration-independent ion sizes.
}%
The importance of the dielectric decrement on the activity coefficient,
or equivalently, the excess chemical potential of electrolyte solutions,
was pointed out in the framework of the mean spherical
approximation~(MSA)~\cite{Fawcett2004Liquids}.  Using the experimentally
measured dielectric constant, theoretical model combined with
Monte-Carlo simulations was
developed~\cite{Vincze2010Nonmonotonic,Valisko2014Effect}.

Theoretical approaches to explain nonlinear dielectric decrement have
been suggested and are based on the dielectric profile inside the
hydration shell, and the interactions between solvent dipoles and ionic
charges~\cite{Levy2012Dielectric,Levy2013Dipolar,MariboMogensen2013Modeling,Duan2015New}.
By considering the free-energy of the solution as a function of the
ionic and dipolar degrees of freedom, the fluctuations in the ion and dipolar-solvent concentrations are taken into account by field-theoretical
calculations of one-loop expansion beyond mean-field
theory~\cite{Levy2012Dielectric,Levy2013Dipolar}.  The obtained dielectric constant is in qualitative agreement with the
general trend of nonlinear behavior at high \(n_{b}\) using a single
fitting parameter related to the dipolar and ionic size.
\changes{However, the basis of this model does not include salt-specific effects. The single minimum cutoff length parameter in this model is assumed to be independent of salt concentration.}

Another approach~\cite{MariboMogensen2013Modeling,Duan2015New} uses the
Booth model~\cite{Booth1951Dielectric,Booth1955Dielectric}, which
accounts for the reorganization of solvent dipoles induced by the
interactions between dipoles and the electric field around ions.  The
resulting prediction~\cite{MariboMogensen2013Modeling} and Monte-Carlo
simulations~\cite{Duan2015New} give a qualitative agreement with the
experimental data.  From these approaches, we deduce that the dielectric
decrement is primarily due to the hydration shell structure around the
ions.  However, it is still not clear what are the important physical
and chemical properties of the ions and solvent, which determine the
salt-specific behavior of the nonlinear decrement at high \(n_{b}\).

In concentrated solutions 
\changes{and locally concentrated region in electric double layer}, hydration shells around ions are in contact with
each other and even overlap, although the salt concentrations are far
below the solubility limit. In other words, the ions in concentrated
solutions are partially dehydrated compared to the full hydration at
dilute solutions.
The degree of dehydration should change with \(n_{b}\) and depends on
the type of ions.
Due to the partial dehydration, dielectric decrement by individual ions
should be weaker than that in dilute solutions.
To elaborate the observed nonlinear dielectric decrement at high
\(n_{b}\), the effect of partial dehydration on the variation of
\(\varepsilon\) should be discussed.

In the present work, we investigate the nonlinear dielectric decrement
at high salt concentrations.  
In Sec.~\ref{sec:models}, 
based on Bruggeman--Hanai's effective medium model
for the dielectric constant of heterogeneous
media~\cite{Bruggeman1935Berechnung,Hanai1960Theory},
we present equations which relate the \(n_{b}\)-dependence of
\(\varepsilon\) to the properties of hydrated ions.
In Sec.~\ref{sec:results_and_discussion}, 
analysis of monovalent salt data suggests that partial dehydration at
high salinity can have a significant contribution to the nonlinear
behavior of \(\varepsilon(n_{b})\). The tendency of dehydration is found
to be related to the solvation free energy and is salt-specific.

\section{Theory}
\label{sec:models}
Dielectric constants of aqueous electrolyte solutions are mainly
described by polarizabilities of water medium and hydrated ions.
In dilute limit where the net dielectric constant \(\varepsilon\)
linearly depends on the ionic concentrations, \(n_{\pm}\), as
\(\varepsilon=\varepsilon_{w}-\gamma_{+}n_{+}-\gamma_{-}n_{-}\), 
the coefficients of dielectric decrement by single ions according to
Clausius-Mossotti relation are given by
\begin{align}
\label{eq:gamma}
 \gamma_{\pm}
&=
3v_{\pm}\varepsilon_{w}
\frac{\varepsilon_{w}-\varepsilon_{\pm}}{2\varepsilon_{w}+\varepsilon_{\pm}}\,,
\end{align}
where \(\varepsilon_{w}\) is the dielectric constant of water solvent,
and \(\varepsilon_{\pm}\) and \(v_{\pm}\) are respectively
the dielectric constant and the volume of hydrated cation (anion).
Due to reduced orientational polarizability of the water molecules within the
hydration shell, the relation \(\varepsilon_{\pm}<\varepsilon_{w}\) holds leading to
\(\gamma_{\pm}>0\).

For finite small volume fractions of the ions, \(\varepsilon\) is
predicted based on Maxwell Garnett model~\cite{Maxwell1904Colours} by
\begin{align}
\label{eq:maxwell--garnett}
\frac{\varepsilon_{w}-\varepsilon}{2\varepsilon_{w}+\varepsilon}
&=\frac{1}{3\varepsilon_{w}}\sum_{\alpha=\pm}\gamma_{\alpha}n_{\alpha}\,.
\end{align}
Since Eq.~(\ref{eq:maxwell--garnett}) still assumes the linear 
superposition of the contributions from dielectric holes of 
hydrated ions, its applicability is limited to relatively low 
concentrations, and was reported at most the volume fraction of 
0.4 although Eq.~(\ref{eq:maxwell--garnett}) predicts some nonlinearity of \(\varepsilon(n_b)\)~\cite{Sareni1996Effective}.
As an example, the volume fraction of 0.4 for alkali halides
corresponds to the salt concentrations about 1.7 to 2.2\,M.
In fact, experimental \(\varepsilon\) for various monovalent electrolyte
solutions
follows Eq.~(\ref{eq:maxwell--garnett}) up to the salt concentration \(n_{b}\approx 2\)\,M, whereas at
higher \(n_{b}\) the weaker decrement than predicted by
Eq.~(\ref{eq:maxwell--garnett}) is observed.

Nevertheless, since 
\(\varepsilon\) by Maxwell Garnett model shows a type of nonlinear
\(n_{b}\)-dependence, 
nonlinear dielectric decrement was analyzed by Maxwell Garnett
model~\cite{Wei1990Dielectric,Biesheuvel2005Volume,LopezGarcia2014Influence}.
Using Maxwell Garnett model, 
experimental \(\varepsilon\) of aqueous LiCl solution was
analyzed~\cite{Wei1990Dielectric}, and dielectric decrement by
ions~\cite{LopezGarcia2014Influence} and
polyelectrolytes~\cite{Biesheuvel2005Volume} in the electric double
layer were modeled.

To predict dielectric responses of concentrated systems, 
Bruggeman developed an effective medium 
theory~\cite{Bruggeman1935Berechnung}, which was later extended 
to the frequency domain by Hanai~\cite{Hanai1960Theory,vanBeek1967Dielectric}.
In the theory, each inclusion is considered to be dispersed in an
effective medium of the dielectric constant \(\varepsilon\) at a
finite volume fraction \(\phi\). 
The applicability of Bruggeman--Hanai model was reported up to 
the volume fraction of 0.8~\cite{Asami2002Characterization}.
In this section, we extend Bruggeman--Hanai equation to the case of
hydrated ionic solutions to analyze the nonlinear dielectric decrement
at high salt concentrations.

\subsection{Fully-hydrated electrolyte solutions} 
We consider symmetric electrolyte solutions of the salt concentration
\(n_{b}\).
The concentrations of dissociated cation and anion are
\(n_{\pm}=n_{b}\).
By applying the effective medium theory to this ionic solution based on
the volumes of fully-hydrated ions, \(v_{\pm}\), 
a differential equation for \(\varepsilon(n_{b})\) is derived as (see derivation in 
\ref{sec:effective_medium})
\begin{align}
\label{eq:bh1}
\frac{\upd \varepsilon}{\upd n_{b}} 
&=
\frac{3\varepsilon}{
1-\left(v_{+}+v_{-}\right) n_{b}
}
\sum_{\alpha=\pm}
\frac{\varepsilon_{\alpha}-\varepsilon}{
\varepsilon_{\alpha}+2\varepsilon
}
v_{\alpha}\,.
\end{align}
The dielectric constants of hydrated cation and anion,
\(\varepsilon_{\pm}\), can be obtained
by inserting the experimentally determined \(\gamma_{\pm}\) and \(v_{\pm}\) into Eq.~(\ref{eq:gamma}).
By solving Eq.~(\ref{eq:bh1}) from \(\varepsilon(n_{b}=
0)=\varepsilon_{w}\), the variation of \(\varepsilon(n_{b})\) is obtained.
If \(\varepsilon_{\pm}<\varepsilon\) holds, \(\varepsilon\) decreases
with \(n_{b}\), and this is usually the case for monovalent electrolytes.
In general, Eq.~(\ref{eq:bh1}) predicts nonlinear dependence of
\(\varepsilon\) on \(n_{b}\).
For later convenience, we define \(n_b\)-dependent coefficient of dielectric 
decrement as
\begin{align}
\label{eq:nbdepgamma1}
 \gamma_{\alpha}(n_{b}) &= 
3v_{\alpha}
\varepsilon(n_{b})\frac{\varepsilon(n_{b})-\varepsilon_{\alpha}}{2\varepsilon(n_{b})+\varepsilon_{\alpha}}.
\end{align}
Using \(\gamma_{\pm}(n_{b})\), the differential 
equation~(\ref{eq:bh1}) is expressed as
\begin{align}
\label{eq:bh1a}
\frac{\upd \varepsilon}{\upd n_{b}} 
&=
\frac{-\sum_{\alpha=\pm}\gamma_{\alpha}(n_{b})}{
1-\left(v_{+}+v_{-}\right) n_{b}
}.
\end{align}

We note that the original Bruggeman--Hanai model was developed 
for a single dispersed component with the closed solution (see
\ref{sec:effective_medium}).
For two-dispersed-component systems, the closed form solution was derived 
by Grosse~\cite{grosse1979extension}.
In this manuscript, the differential form of Bruggeman--Hanai 
model is used to consider the effect of partial dehydration since
the differential form is convenient for considering 
\(n_{b}\)-dependent modifications.

\subsection{Partial dehydration at high-salt concentrations} 
The volume fraction of the dissociated ions is defined as
\(\phi=\left(v_{+}+v_{-}\right)n_{b}\).
For monovalent small ions, a typical value of hydration radius is
0.33-0.38\,nm~\cite{Nightingale1959Phenomenological}.
Assuming the cations and anions are fully hydrated even at high salt
concentrations, 
maximal \(n_{b}\) lies at 3.5-5.5\,M, which is typically
far below the solubility concentrations.
This fact indicates that the ions in concentrated solutions are partially
dehydrated.
In other words, some water molecules forming hydration shell in 
dilute solutions are partly unbounded in concentrated solutions.
The effective volume of a partially hydrated ion is smaller than 
that of a fully hydrated ion.
Since Eq.~(\ref{eq:bh1}) assumes the volume of fully-hydrated ions, it
cannot be applied to high-\(n_{b}\) region where the effects of partial
dehydration become substantial.

When the partial dehydration occurs, the hydration volumes are 
reduced, which leads that dielectric decrement becomes weaker as 
demonstrated in Eq.~(\ref{eq:nbdepgamma1}).
In order to discuss 
the nonlinear dielectric decrement at high-\(n_{b}\), we modify 
Eq.~(\ref{eq:bh1}) to take into account the partial dehydration.
Besides the volume reduction of the hydration ions, 
several other physical effects are possible
when partial dehydration of ions in a concentrated solution 
occurs. 
Since the hydration shell screens the ionic charge, partial 
dehydration enhances Coulombic interactions among ions.
As a result, ion-pair formation tendency becomes stronger due to 
the partial dehydration in addition to the decrease of inter-ion 
distance with an increase of the salt concentration.
The formation of ion pairs might modify the dielectric response 
of a solution.
In either case with or without ion pair formation, the 
dielectric decrement by partially dehydrated ions is 
qualitatively weaker than that by fully hydrated ions.
Enhanced Coulombic interaction would also modify the hydration 
shell permittivity.
Moreover, once partial dehydration occurs, the fraction of 
partially dehydrated ions may not be uniform in the solution. 
The collectivity of ions in the Kirkwood correlation length is not clear
for concentrated solutions~\cite{zhang2016computing}.
As such, the possible effects induced by partial dehydration can 
be complicated. 
However, it is not clear whether these phenomena have a 
substantial effect on the solution dielectric decrement.
In this paper, we focus on the volume reduction of hydration 
shell as a primary effect on solution dielectric decrement by 
partial dehydration.

The variation of \(v_{\pm}\) should be directly related to the 
flexibility of the hydration shell, which reflects how strong an 
ion can associate with water molecules over water-water 
interaction.  In addition, interaction between cation and anion 
can affect dehydration as \(n_{b}\) is larger where inter-ion 
distance is decreased.  
In general, identifying 
\(n_{b}\)-dependence of \(v_{\pm}\) for various ions is a 
complicated task.
Evaluation of the solution dielectric constant from the 
orientaional correlation through Monte-Carlo or molecular 
dynamics simulations is rather difficult~\cite{zhang2016computing}.
In 
what follows, we focus on the effect of the partial dehydration 
on the nonlinear dielectric decrement by considering a simplified 
model of ion-specific dehydration behaviors.

We assume that partial dehydration starts when the volume 
fraction of the ions reaches a certain threshold value, 
\(\phi_{p}\). For \(\phi>\phi_{p}\), the volumes of hydrated 
cations and anions decrease as \(f(n_{b})v_{\pm}\) with a factor 
\(f<1\)
due to the partial loss of 
hydration shell. The modified equation for \(\varepsilon(n_{b})\) 
reads
\begin{align}
\label{eq:bh2}
\frac{\upd \varepsilon}{\upd n_{b}} 
&=
\frac{-\sum_{\alpha=\pm}\tilde{\gamma}_{\alpha}(n_{b})}{
1-f (n_{b})\left(v_{+}+v_{-}\right) n_{b}
}
,
\end{align}
where the \(n_b\)-dependent coefficient of dielectric 
decrement is modified as. 
\begin{align}
\label{eq:nbdepgamma2}
  \tilde{\gamma}_{\alpha}(n_{b}) &= 
3f(n_{b})v_{\alpha}
\varepsilon(n_{b})\frac{\varepsilon(n_{b})-\varepsilon_{\alpha}}{2\varepsilon(n_{b})+\varepsilon_{\alpha}}.
\end{align}
This modified decrement coefficient (\ref{eq:nbdepgamma2}) 
becomes smaller as a result of decreasing volume of partially 
hydrated ions.
Considering the volume reduction of hydration shell
not only resolves the inconsistency between the observed 
solubility and the assumption of fully hydrated ions, 
but also weakens the dielectric decrement coefficient at high salt 
concentrations.
The onset volume fraction, \(\phi_{p}\), depends on the salt. Larger
\(\phi_{p}\) indicates the smaller tendency of ions to be dehydrated,
namely, the ions strongly associate with water molecules. Conversely,
ions that weakly associate with water would give smaller \(\phi_{p}\).
To proceed further, we assume an explicit form of \(f(n_{b})\) 
just to fit the experimental data by
\begin{align}
\label{eq:phip_definition}
f(n_{b}) &=  
\begin{cases}
1 & \phi<\phi_{p}\,,
\\
\frac{\phi_{p}}{\left(v_{+}+v{-}\right)n_{b}} & \text{otherwise}\,.
\end{cases}
\end{align}
In this model of \(f\), we simply consider the decrease of the hydration
 volumes with \(n_{b}\) after the onset of the partial dehydration.
More realistic forms of the hydration volumes will not be so simple as
Eq.~(\ref{eq:phip_definition}). 
Nonetheless, the model of Eqs.~(\ref{eq:bh2})
and~(\ref{eq:phip_definition}) can be applied around the onset of
partial dehydration and weakened dielectric decrement 
\changes{and thus is used to extract the salt-dependent onset volume fraction of the partial dehydration.}

\section{Results and Discussion}
\label{sec:results_and_discussion} 
Experimental data of the dielectric constant of symmetric 1:1
electrolyte solutions at \(T=25^{\circ}\)C are taken from
Ref.~\citenum{Barthel1995Electrolyte}.  
To make sure that the dielectric constant data are those of the solutions,
solubility data of some
electrolytes in water solvent from 
Ref.~\citenum{Perry2011Handbook}
is shown in units of M in Table~\ref{tbl:solubility}.  
In addition, volume fractions of dissociated ions at the 
solubility salt concentration, \(n_{b}^{s}\), are estimated based 
on two types of ion sizes. The volume fraction based on the 
crystallographic radius~\cite{Shannon1969Effective}, \(\phi_{c}\), 
and that based on the fully-hydrated ion 
size~\cite{Nightingale1959Phenomenological}, \(\phi_{h}\), are 
also shown in Table~\ref{tbl:solubility}.
By definition, since a hydrated ion size is larger than the bare 
ion size, \(\phi_c<\phi_{h}\) holds.
The value of \(\phi_{h}\) spreads from 0.77 to 3.25, which fact
reflects ion-dependent water affinity. 
For all the salt except KCl in Table~\ref{tbl:solubility}, 
\(\phi_{h}\) exceeds unity,  which fact leads that ions dissolved 
in concentrated solutions do not maintain the fully hydrated 
structure formed in the dilute solutions, and thus the ions are 
dehydrated to some extent in concentrated solutions.
In contrast, the value of \(\phi_{c}\) does not vary so much as
\(\phi_{h}\) and is about from 0.1 to 0.2. 
In this range of volume fraction, almost bare cations and anions can closely
approach to each other. This fact indicates that \(\phi_{c}\) might be
utilized to predict the solubility of electrolytes.

\begin{table}[htbp]
\caption{\label{tbl:solubility} Solubility concentration, \(n_{b}^{s}\),
of various salts in water solvent.  Corresponding volume fraction based
 on the crystallographic radius~\cite{Shannon1969Effective} is denoted
 by \(\phi_{c}\) and that based on the fully-hydrated ion
 size~\cite{Nightingale1959Phenomenological} is denoted by \(\phi_{h}\).
 }
\begin{tabular}{lccc}
\hline
\hline
 & \(n_{b}^{s}\) [M] & \multicolumn{2}{c}{Volume fraction} \\
\cline{3-4}
Salt & & \(\phi_{c}\) & \(\phi_{h}\) \\
\hline
 LiCl & 13.9 & 0.188 & 3.25 \\
 NaCl & 5.41 & 0.085 & 1.13 \\
 KCl  & 4.17 & 0.086 & 0.77 \\
 RbCl & 5.80 & 0.131 & 1.06 \\
 CsCl & 6.69 & 0.184 & 1.22 \\
 KF   & 13.2 & 0.174 & 2.67 \\
 KI   & 5.88 & 0.182 & 1.08 \\
 LiBr & 12.2 & 0.207 & 2.83 \\
\hline\hline
\end{tabular}
\end{table}

\subsection{Fully-hydrated electrolyte models} 
We first discuss the applicability
of 
fully-hydrated electrolyte models that are
Maxwell Garnett model of Eq.~(\ref{eq:maxwell--garnett}) and
Bruggeman--Hanai model of Eq.~(\ref{eq:bh1}).
Parameters of hydrated ions are shown in Table~\ref{tbl:ion_parameters}.
The hydration radius is taken from
Ref.~\cite{Nightingale1959Phenomenological}.
The coefficient of linear dielectric decrement, \(\gamma_{\pm}\), is
determined from the experimental data of \(\varepsilon\) by fitting the
linear relationship
\(\varepsilon=\varepsilon_{w}-\left(\gamma_{+}+\gamma_{-}\right)n_{b}\)
at \(n_{b}<2\)\,M.
The (\(\gamma_{+}+\gamma_{-}\)) from different salts is decomposed to ionic
\(\gamma_{\pm}\) by the least squares method by setting a reference value of
\(\gamma_{\text{Na}^{+}}=8\) as was done in
Ref.~\cite{Hasted1948Dielectric}.
By substituting \(\gamma_{\pm}\), \(v_{\pm}\), and \(\varepsilon_{w}=78.3\)
into Eq.~(\ref{eq:gamma}), \(\varepsilon_{\pm}\) is obtained.

\begin{table}[htbp]
\caption{\label{tbl:ion_parameters}
Coefficients of linear dielectric decrement, \(\gamma_{\pm}\) estimated from
 experimental \(\varepsilon(n_{b})\) taken from
 Ref.~\cite{Barthel1995Electrolyte}, and experimentally obtained hydration radius~\cite{Nightingale1959Phenomenological}, $r_{\text{h}}$,
for several monovalent cations and anions.
}
\begin{tabular}{ccc|ccc}
\hline
\hline
   & 
  \multicolumn{1}{c}{
\text{\(\gamma\)\,[M\(^{-1}\)]}
} 
& 
  \multicolumn{1}{c|}{
\text{$r_{\text{h}}$\,[nm]}
} 
&
   & 
  \multicolumn{1}{c}{
\text{\(\gamma\)\,[M\(^{-1}\)]}
} 
& 
  \multicolumn{1}{c}{
\text{$r_{\text{h}}$\,[nm]}
} 
\\
  \hline
Li$^{+}$  & 9.67 & 0.382
&
F$^{-}$  & 1.83 & 0.352
\\
Na$^{+}$  & 8 & 0.358
&
Cl$^{-}$  & 3.95 & 0.332
\\
K$^{+}$  & 7.19 & 0.331
&
I$^{-}$  & 4.16 & 0.331
 \\
Rb$^{+}$  & 7.99 & 0.329
&
Br$^{-}$  & 4.51 & 0.330
\\
Cs$^{+}$  & 6.44 & 0.329
&
NO\(_{3}^{-}\)  & 3.68 & 0.335
\\
Et\(_{4}\)N$^{+}$  & 14.2 & 0.400
&
ClO$_{4}^{-}$  & 5.11 & 0.338
\\
\hline
\hline
 \end{tabular}
\end{table}

\begin{figure}[htbp]
 \centering
\includegraphics[width=83mm]{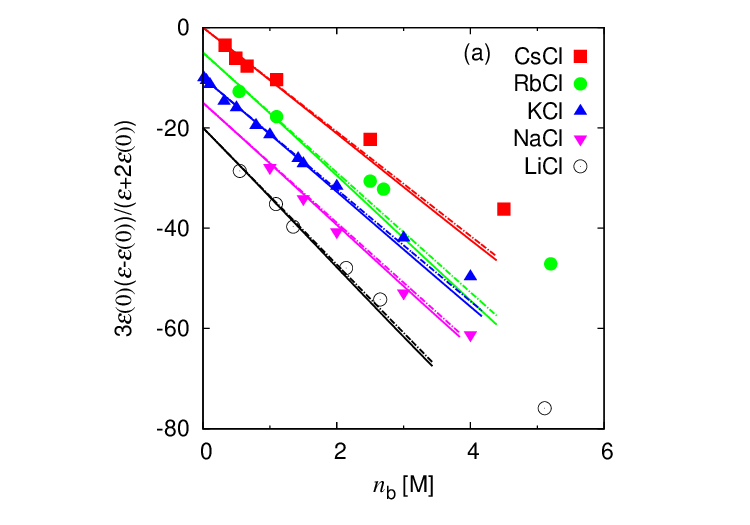}
\\
\includegraphics[width=83mm]{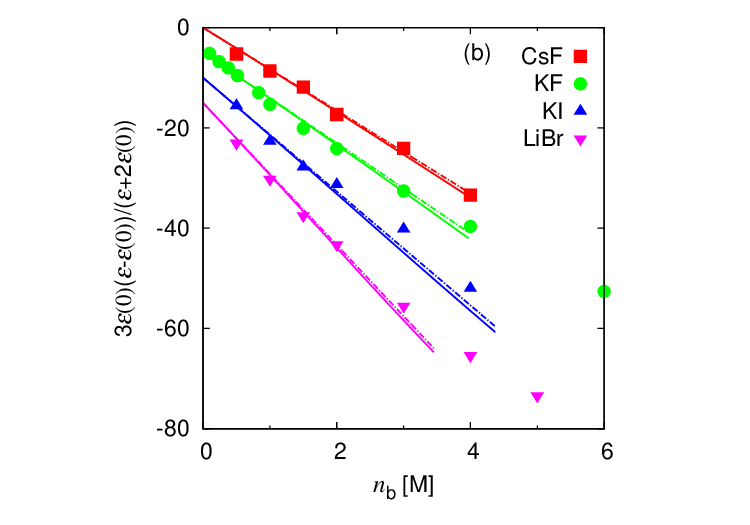}
\\
\includegraphics[width=83mm]{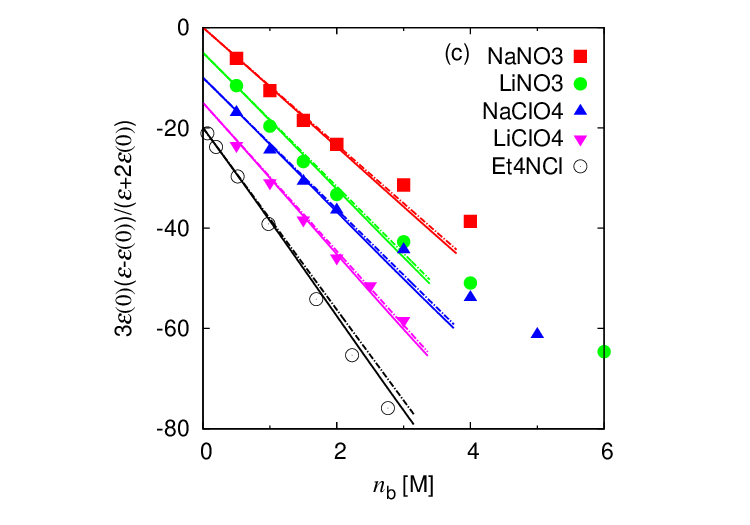}
\caption{\textsf{
\label{fig1}
Comparison of the 
Maxwell Garnett model of Eq.~(\ref{eq:maxwell--garnett}) 
 (dash-dotted line) and the fully-hydrated electrolyte model of Eq.~(\ref{eq:bh1}) (solid line)
with 
experimental data from
Ref.~\citenum{Barthel1995Electrolyte} for various salts.
To check the linearity assumption in Maxwell Garnett model, 
\(3\varepsilon(0)\left[\varepsilon(n_{b})-\varepsilon(0)\right]/\left[\varepsilon(n_{b})+2\varepsilon(0)\right]\) as a function of 
salt concentration \(n_{b}\) is plotted.
The values of 
\(\varepsilon\) are shifted for clarity purpose only.
(a) 
CsCl (red square), 
RbCl (green circle, -5), 
KCl (blue upper triangle, -10), 
NaCl (pink lower triangle, -15), 
LiCl (black open circle, -20), 
(b) 
CsF (red square), 
KF (green circle, -5), 
KI (blue upper triangle, -10), 
LiBr (pink lower triangle, -15), 
(c) 
NaNO\(_{3}\) (red square), 
LiNO\(_{3}\) (green circle, -5), 
NaClO\(_{4}\) (blue upper triangle, -10), 
LiClO\(_{4}\) (pink lower triangle, -15), 
Et\(_{4}\)NCl
(black open circle, -20). 
}}
\end{figure}

In Fig.~\ref{fig1}, we compare the Maxwell Garnett model of 
Eq.~(\ref{eq:maxwell--garnett}) (dash-dotted line)
and
the Bruggeman--Hanai model of Eq.~(\ref{eq:bh1}) solved using a 
fourth order Runge--Kutta method (solid line)
to experimental values for 
fourteen different electrolyte solutions. 
To check the linearity assumption in Maxwell Garnett model, 
\(3\varepsilon(0)\left[\varepsilon(n_{b})-\varepsilon(0)\right]/\left[\varepsilon(n_{b})+2\varepsilon(0)\right]\) as a function of 
salt concentration \(n_{b}\) is plotted in Fig.~\ref{fig1}.
For each salt, the models are calculated up to the volume 
fraction of 0.8, and thus the model functions by 
Eqs.~(\ref{eq:maxwell--garnett}) and (\ref{eq:bh1}) end at around 
\(n_{b}=3.1-4.4\) for the salts in Fig.~\ref{fig1}, which 
demonstrates that the fully-hydrated electrolyte models based on 
the hydration volume at dilute conditions do not 
extend to the solubility concentration.
Comparing the two models, there is almost no 
difference between them
although the dielectric constant by Bruggeman--Hanai model 
is slightly smaller than that by Maxwell Garnett model.
Although the comparison can be made in a limited range of 
\(n_{b}\) for each salt due to the free hydrated ion 
volume, \(\varepsilon\) by both models describes the observed 
dielectric decrement up to 2-3\,M, and the range of applicability 
of Eqs.~(\ref{eq:maxwell--garnett}) and~(\ref{eq:bh1}) is 
salt-specific.
For higher \(n_{b}\), experimental \(\varepsilon\) shows weaker 
decrement than predicted by the fully-hydrated electrolyte models. 
This observation suggests that the effects of partial dehydration 
should be considered in order to explain the variation of 
\(\varepsilon\) observed in high-\(n_{b}\) range.

\subsection{Partially-dehydrated electrolyte model} 

\begin{figure}[htbp]
 \centering
\vspace*{-10ex}
\includegraphics[width=83mm]{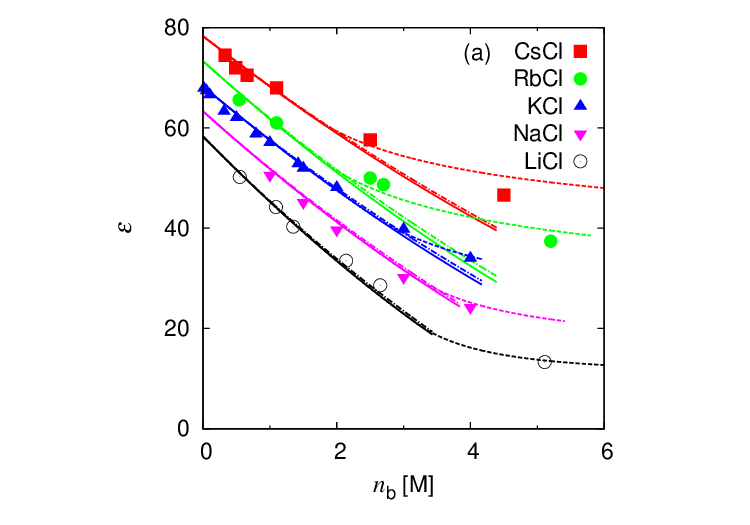}
\\
\includegraphics[width=83mm]{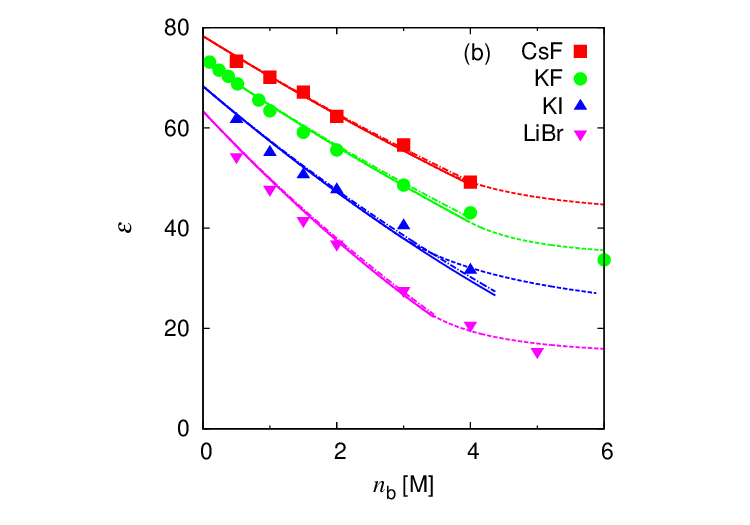}
\\
\includegraphics[width=83mm]{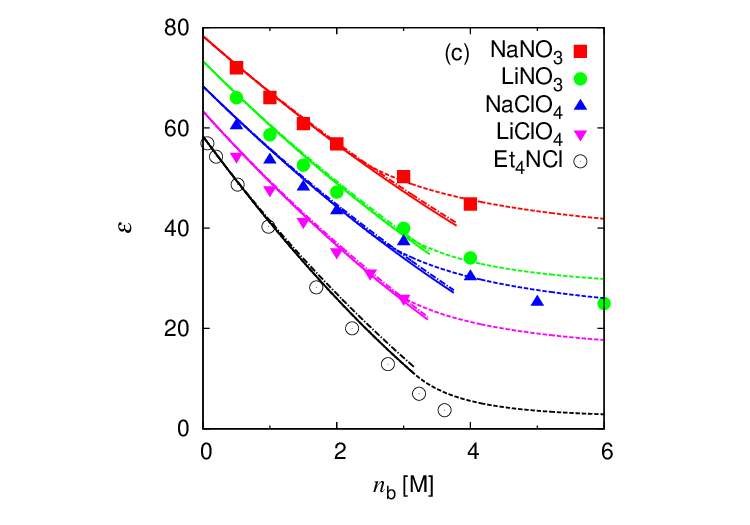}
 \caption{\textsf{
\label{fig2}
Comparison of the dielectric constants from the
the partial dehydration model of Eqs.~(\ref{eq:bh2}) and
 (\ref{eq:phip_definition})~(dashed line), the full hydration 
 model of Eq.~(\ref{eq:bh1})~(solid line), and Maxwell Garnett 
 model of Eq.~(\ref{eq:maxwell--garnett})~(dash-dotted line), with
 experimental data from 
Ref.~\citenum{Barthel1995Electrolyte}~(symbol),
as a function of salt concentration \(n_{b}\) for various salts. 
For the salts whose solubility, \(n_{b}^{s}\), is known, the solutions of
 Eq.~(\ref{eq:bh2}) are up to \(n_{b}^{s}\), otherwise
 Eq.~(\ref{eq:bh2}) is solved up to \(n_{b}=6\)\,M.
The solutions of the full hydration model and Maxwell Garnett 
 model are for \(\left(v_{+}+v_{-}\right)n_{b}<0.8\). 
The values of \(\varepsilon\) are shifted for 
clarity purpose only: in (a) 
CsCl (red square), 
RbCl (green circle, -5), 
KCl (blue upper triangle and dotted blue line, -10),
NaCl (pink lower triangle and dash-dotted pink line, -15),
LiCl (black open circle and dash-dot-dotted black line, -20);
in (b) 
CsF (red square and solid red line), 
KF (green circle and dashed green line, -5), 
KI (blue upper triangle and dotted blue line, -10),
LiBr (pink lower triangle and dash-dotted pink line, -15);
in (c) 
NaNO\(_{3}\) (red square and solid red line), 
LiNO\(_{3}\) (green circle and dashed green line, -5), 
NaClO\(_{4}\) (blue upper triangle and dotted blue line, -10),
LiClO\(_{4}\) (pink lower triangle and dash-dotted pink line, -15),
Et\(_{4}\)NCl
(black open circle and dash-dot-dotted black line, -20).
}}
\end{figure}

\begin{figure}[htbp]
 \centering
\includegraphics[width=\hsize]{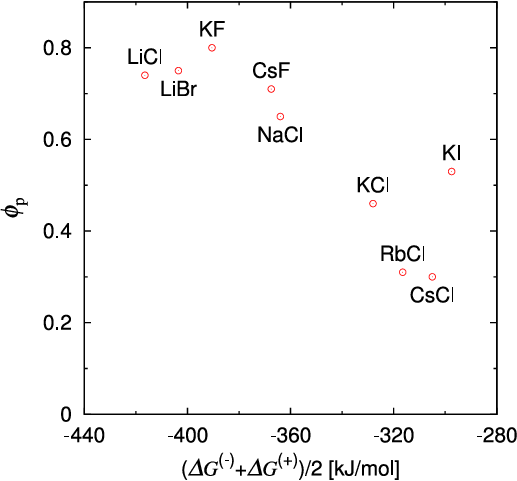}
\caption{
\textsf{ 
Volume fraction at the onset of partial dehydration, \(\phi_{p}\), in the partial
 dehydration model of Eq.~(\ref{eq:bh2}), plotted against the mean
solvation free energy of cation and anion~\cite{Fawcett1999Thermodynamic,Fawcett2004Liquids}.
}}
\label{fig3}
\end{figure}

Next, we examine the model with partial dehydration of
Eqs.~(\ref{eq:bh2}) and (\ref{eq:phip_definition}).
In Fig.~\ref{fig2}, the dielectric constant by Eq.~(\ref{eq:bh2}) is
compared to the experimental values.
Eq.~(\ref{eq:bh2}) is solved up to the solubility concentration if it
is available, otherwise up to \(n_{b}<6\)\,M (dashed line).
For reference, the predictions by fully-hydrated electrolyte 
models~(Maxwell Garnett and Bruggeman--Hanai models) are also 
drawn in Fig.~\ref{fig2}.
The dielectric constant by the partial dehydration model in 
Fig.~\ref{fig2} is obtained by fitting to the experimental data 
through the onset of partial dehydration, \(\phi_{p}\), while 
keeping the other parameters, \(\varepsilon_{\pm}\) and 
\(v_{\pm}\) fixed to the values in Table~\ref{tbl:ion_parameters}. 
The obtained value of \(\phi_{p}\) is plotted in Fig.~\ref{fig3}.
Due to
the partial dehydration, the available solution of
Eq.~(\ref{eq:bh2}) is extended to higher-\(n_{b}\) range than that of
the fully-hydrated electrolyte model of Eq.~(\ref{eq:bh1}), and weakened
dielectric decrement at high-\(n_{b}\) values is successfully reproduced
up to \(n_{b}<6\)\,M.

The obtained value of \(\phi_{p}\) in Fig.~\ref{fig3} depends on the type of salt, which fact
is supposed that \(\phi_{p}\) reflects the tendency of dehydration of
ions.
The dehydration tendency can be measured by the solvation free energy of
cation and anion. The larger the magnitude of solvation free energy is,
the stronger the ion should associate with water molecules leading
the smaller tendency of dehydration.
In Fig.~\ref{fig3}, we plot the parameter \(\phi_{p}\) against the mean
solvation free energy of cation and anion taken from
Refs.~\citenum{Fawcett1999Thermodynamic,Fawcett2004Liquids}.
We observe some trend that \(\phi_{p}\) becomes larger for larger
\(\left|\Delta G^{(+)}+\Delta G^{(-)}\right|/2\). 
This fact suggests that 
\(\phi_{p}\), determined from \(\varepsilon\) is qualitatively related to the tendency of dehydration of different salts.

Nonlinear variation of \(\varepsilon\) in Fig.~\ref{fig2} is 
classified into at least two regimes. For smaller \(n_{b}\), 
variation of \(\varepsilon\) is determined by the fully-hydrated 
ions by 
Eqs.~(\ref{eq:maxwell--garnett}) or
 (\ref{eq:bh1}), while 
deviation from 
Eqs.~(\ref{eq:maxwell--garnett}) or
(\ref{eq:bh1}) and weakened decrement starts 
at a certain value of \(n_{b}\). The changeover between them is 
related to the mean solvation free energy of the cation and the 
anion. For example, in Fig.~\ref{fig2}(a), the changeover 
\(n_{b}\) value to weakened decrement is smaller for CsCl and 
RbCl with smaller solvation free energy, than for NaCl and LiCl 
with larger solvation free energy.

The partial dehydration model of Eq.~(\ref{eq:phip_definition}) works
well around the changeover \(n_{b}\), whereas it is too simplified to be
applied to the higher-\(n_{b}\) variation of \(\varepsilon\).  By the
change of the hydration volume of Eq.~(\ref{eq:phip_definition}),
\(\varepsilon\) from Eq.~(\ref{eq:bh2}) shows a rapid saturation of
\(\varepsilon\) value for \(\phi\gg\phi_{p}\).  
In order to predict the nonlinear variation of \(\varepsilon\) up 
to the solubility concentration, more realistic model for partial 
dehydration than Eq.~(\ref{eq:phip_definition}) is required. 
For this modelling, it is necessary to analyze the hydration 
shell structure in highly concentrated solution, the distribution 
of partially hydrated ions, and the dielectric response of 
individual hydrated ions and the solution.
This challenging task is an important future issue.

\section{Conclusions}
We have presented a model exploring the salt-specific nonlinear 
behavior of \(\varepsilon(n_{b})\) of the dielectric constant of 
electrolyte solution as 
function of the salt concentration, \(n_{b}\)~\cite{Barthel1995Electrolyte,Glueckauf1964Bulk,Wei1990Dielectric,Wei1992Ion}.
To explain the nonlinear dielectric decrement as observed in 
experiment~\cite{Barthel1995Electrolyte,Glueckauf1964Bulk,Wei1990Dielectric,Wei1992Ion}, 
we considered
the effect of partial dehydration in 
concentrated solutions, which is suggested by the 
inconsistency between the observed salt 
solubility~\cite{Perry2011Handbook} and the fully hydrated 
ion volume~\cite{Nightingale1959Phenomenological} in concentrated solutions, and has not been explicitly considered in the previous models~\cite{Levy2012Dielectric,Levy2013Dipolar,MariboMogensen2013Modeling,Duan2015New}.
The partial dehydration effect
is taken into 
account by extending the Bruggeman--Hanai dielectric 
model~\cite{Bruggeman1935Berechnung,Hanai1960Theory,vanBeek1967Dielectric,Asami2002Characterization,grosse1979extension}.
Our model can explain well the experimental nonlinear dielectric
decrement behavior over a wide range of salt concentrations, up to 6\,M
beyond the applicable \(n_{b}\) of the fully hydrated electrolyte models
~\cite{Maxwell1904Colours,Bruggeman1935Berechnung}.
While the small-\(n_{b}\) variation of \(\varepsilon\) is 
described by the dielectric cavities by the fully-hydrated 
cations and anions, weakened dielectric decrement at 
high-\(n_{b}\) is determined by the partial dehydration of the 
ions.
The onset volume fraction of the partial dehydration, which reflects 
the tendency of dehydration of the ions, is found to be 
salt-specific~\cite{BenYaakov2011Dielectric,Kunz2009Specific},
and is consistent with the solvation free energy~\cite{Fawcett1999Thermodynamic,Fawcett2004Liquids}.
Accurate modeling of \(\varepsilon(n_{b})\) is required when 
considering the electrostatic and solvation interactions in 
various applications.
Our model of \(\varepsilon(n_{b})\) presents a step forward a better
understanding of the underlying physical principles, 
especially in systems of high ionic concentrations, such as nanofluidic
devices, electrokinetic phenomena near high-voltage surfaces, and
crowding effects in biological cells.
One advantage of our model is that it accounts the separate
contributions from cations and anions to
\(\varepsilon=\varepsilon(n_{+},n_{-})\). 
Hence, not only it describes
\(\varepsilon(n_{b})\) in the bulk, but it can be directly applicable to
inhomogeneous and local environments, such as those formed in electric
double layers~\cite{BenYaakov2011Dielectric,Biesheuvel2005Volume,LopezGarcia2014Influence,Hatlo2012Electric,Nakayama2015Differential,Gillespie2015Review,gupta2018electrical,lesniewska2023electrostatics}
 and in confined nanochannels, which are highly relevant to colloidal transport phenomena~\cite{saha2022shielding} and interactions between colloids and interfaces~\cite{gao2023effects}.
In this paper, the parameters in our model were obtained from the
fitting to experimental data. 
The determination of the partial
dehydration behavior from the molecular level~\cite{zhang2016computing} remains an important issue for future studies.

\changes{
Lastly, we comment on the applicability of our model for 
asymmetric and/or multivalent electrolytes.
The reported dielectric constants for multivalent ionic solutions 
show qualitatively different dependence on the salt concentration 
from those for symmetric monovalent 
electrolytes~\cite{Barthel1995Electrolyte}.
Depending on the salt type, the dielectric constant for 
multivalent ions may even increase or decrease with salt 
concentration.
Therefore, the partial dehydration discussed in this manuscript 
cannot explain all of the dielectric behaviors in multivalent 
cases. This suggests other physics as well as the partial 
dehydration are required for multivalent cases.
One of the important effects in the multivalent cases is the 
contribution of ion association~\cite{Kunz2009Specific}.
Theoretically and experimentally, it is not clear how associated 
ion pairs affect the dielectric response and should be taken into 
account to explore the dielectric behavior in the multivalent 
cases.
}

\appendix

\section{
Effective medium models}
\label{sec:effective_medium}
\subsection{Maxwell Garnett model for dilute systems}
Maxwell Garnett model describes the permittivity of dilute suspensions
where spherical inclusions of a permittivity \(\varepsilon_{h}\) are
immersed in a medium of a permittivity \(\varepsilon_{w}\). The
permittivity of the mixture, \(\varepsilon\), according to 
Maxwell Garnett model is given
by~\cite{Maxwell1904Colours}
\begin{align}
\frac{\varepsilon_{w}-\varepsilon}{2\varepsilon_{w}+\varepsilon}
&=
\frac{\varepsilon_{w}-\varepsilon_{h}}{2\varepsilon_{w}+\varepsilon_{h}}
\phi\,,
\end{align}
where \(\phi\) is the volume fraction of the inclusions.
For electrolyte solutions, hydrated cations and anions work as
dielectric inclusions. Summing up the  contributions from hydrated
cations and anions, Eq.~(\ref{eq:maxwell--garnett}) is obtained.

\subsection{Bruggeman--Hanai model for concentrated suspensions}
When a small amount of the inclusions of the volume \(\upd V\) is added
to a suspension of finite volume fraction \(\phi\) and the volume
\(V\gg\upd V\),
the additional inclusions are supposed as dilute in an
effective medium of the permittivity \(\varepsilon(\phi)\). Therefore,
the change in the permittivity \(\upd \varepsilon\) by \(\upd V\) can be
described by Maxwell Garnett model by~\cite{Bruggeman1935Berechnung,Hanai1960Theory}
\begin{align}
\label{eq:depsilon}
 \frac{
\varepsilon-\left(\varepsilon+\upd \varepsilon\right)
}{
2\varepsilon+\left(\varepsilon+\upd \varepsilon\right)
}
&=
 \frac{\varepsilon-\varepsilon_{h}}{2\varepsilon+\varepsilon_{h}
}
\frac{\upd V}{V+\upd V}\,.
\end{align}
Substituting  the change of the volume fraction \(\upd
\phi=\left(1-\phi\right)\upd V/\left(V+\upd V\right)\) into
Eq.~(\ref{eq:depsilon}) yields 
\begin{align}
\label{eq:bh0}
  \frac{
\upd \varepsilon
}{
\upd \phi
}
&=
 \frac{\varepsilon_{h}-\varepsilon}{\varepsilon_{h}+2\varepsilon
}
\frac{3\varepsilon}{1-\phi}\,.
\end{align}
Equation~(\ref{eq:bh0}) is analytically solved with a boundary
condition \(\varepsilon(0)=\varepsilon_{w}\) as
\begin{align}
\left(
\frac{\varepsilon_{w}}{\varepsilon}
\right)^{1/3}
\frac
{\varepsilon_{h}-\varepsilon}
{\varepsilon_{h}-\varepsilon_{w}}
&=
\left(1-\phi\right)\,,
\end{align}
which is the Bruggeman--Hanai equation for a single dispersed component.

The differential equation for the permittivity~(\ref{eq:bh0}) can 
be extended to the case of two dispersed components.
However, its solution is not obtained analytically.
Without directly integrating the differential equation, 
the closed form for two dispersed components was derived by Grosse
\cite{grosse1979extension}.
Consider two dispersed components A and B.
Supposing that the solvent volume \(V_{w}\) is devied into two 
portions \(V'_{w}\) and \(V''_{w}\) which respectively constitute 
two suspensions of A and B, the Bruggeman--Hanai permittivity of 
the suspension is expressed by 
\begin{align}
\left(
\frac{\varepsilon_{w}}{\varepsilon}
\right)^{1/3}
\frac
{\varepsilon_{A}-\varepsilon}
{\varepsilon_{A}-\varepsilon_{w}}
&=
1-\frac{V_{A}}{V_{A}+V'_{w}}\,,
\\
\left(
\frac{\varepsilon_{w}}{\varepsilon}
\right)^{1/3}
\frac
{\varepsilon_{B}-\varepsilon}
{\varepsilon_{B}-\varepsilon_{w}}
&=
1-\frac{V_{B}}{V_{B}+V''_{w}}\,,
\\
V_{w} &= V'_{w}+V''_{w}\,,
\end{align}
where 
\(\varepsilon_{A}\), 
\(\varepsilon_{B}\) are the permittivities of the two dispersed 
components, and \(V_{A}\) and \(V_{B}\) are the volumes.

\subsection{Bruggeman--Hanai model for symmetric electrolyte solutions}
For symmetric electrolyte solutions, increment of a small amount of
the salt yields increment of dissociated cation and anion, and the volume fraction of the
dissociated ions is related to the salt concentration by
\(\phi=\left(v_{+}+v_{-}\right)n_{b}\). By taking into account the
contributions from the hydrated cation and anion in
Eq.~(\ref{eq:depsilon}), we obtain Eq.~(\ref{eq:bh1}).

\section*{Acknowledgements}
The author thanks David Andelman for discussion at the early stage 
 of this work.
The numerical calculations have been partly carried out using the 
computer facilities at the Research Institute for Information 
Technology, Kyushu University. This work has been supported by 
Grants-in-Aid for Scientific Research (JSPS KAKENHI) under Grant 
No.~JP18K03563.

\end{document}